\documentclass[12pt]{article}

\usepackage[dvips,final]{graphicx}
\usepackage{epsfig}

\usepackage{amsmath}

\begin{document}
\begin{flushright}
TTP03-12 \\
\end{flushright}
\vspace{0.5cm}
\thispagestyle{empty}
\begin{center}

{\Large \bf
QCD Light-Cone Sum Rule  Estimate \\
of Charming Penguin Contributions 
in $B \rightarrow \pi\pi$}
\vskip 1.5true cm

{\large\bf
A.~Khodjamirian\,$^{*)}$,
Th.~Mannel\,, B.~Meli\'c\,$^{**)}$}
\\[1cm]~
\vskip 0.3true cm
{\it Institut f\"ur Theoretische Teilchenphysik, Universit\"at
Karlsruhe,\\  D-76128 Karlsruhe, Germany } \\

\end{center}

%\maketitle

\vspace*{1cm}

\begin{abstract}
\noindent
Employing  the QCD light-cone sum rule approach
we calculate the $B \rightarrow \pi\pi$
hadronic matrix element of the current-current operator 
with $c$ quarks in the penguin topology (``charming penguin''). 
The dominant contribution to the sum rule is 
due to the $c$-quark loop at short distances and is of $O(\alpha_s)$ with respect 
to the factorizable $B\to \pi\pi$ amplitude. The effects of soft gluons 
are suppressed at least by $O(\alpha_s m_b^{-2})$. 
Our result indicates that sizable nonperturbative effects generated by
charming penguins at finite $m_b$
are absent. The same is valid for the penguin contractions of 
the current-current operators with light quarks.
\end{abstract}

\vspace*{\fill}

\noindent $^{*)}${\small \it On leave from
Yerevan Physics Institute, 375036 Yerevan, Armenia} \\
\noindent $^{**)}${\small \it On leave from
Rudjer Bo\v skovi\'c Institute, Zagreb, Croatia} \\

\newpage

%\section{Introduction}

{\bf 1.} Charmless two-body hadronic $B$ decays, such as $B\to \pi\pi$,
are promising sources of information on CP-violation
in the $b$-flavour sector. The task of unfolding   
CKM phases from the decay observables is challenged by our 
limited ability to calculate the relevant hadronic interactions. 
One has to resort to approximate
methods based on the expansion in the inverse 
$b$-quark mass. The long-distance effects
are then parametrized in terms of process-independent characteristics: 
heavy-light form factors, hadronic decay constants and 
light-cone distribution 
amplitudes (DA). For example, in the
QCD factorization approach \cite{BBNS} the exclusive $B$-decay amplitudes  
in  the $m_b \to\infty$ limit are expressed in terms of the factorizable part  
and calculable $O(\alpha_s)$ nonfactorizable corrections.
For phenomenological applications it is important to investigate
the subleading effects in the decay amplitudes suppressed by inverse powers of $m_b$.
Especially interesting are ``soft'' nonfactorizable effects, involving 
low-virtuality  gluons and quarks, not necessarily 
accompanied by an $\alpha_s$-suppression.

Quantitative estimates of nonfactorizable 
contributions, including the power-suppressed ones can be obtained \cite{AK} 
using the method of QCD light-cone sum rules (LCSR) \cite{lcsr}.
In particular, the $O(1/m_b)$ soft-gluon corrections have been 
calculated \cite{AK} for the $B\to \pi\pi$ matrix elements 
with the emission topology.
Furthermore, the LCSR estimate  
of the chromomagnetic dipole operator (gluonic penguin) 
contribution to $B\to \pi\pi$ 
was obtained in \cite{KMU}. In this case the soft-gluon 
contribution, being suppressed as $1/m_b^2$, 
at finite $m_b$ is of the magnitude of the $O(\alpha_s)$ 
hard-gluon part. However, the nonfactorizable corrections are nonuniversal, 
depending on the effective operators and quark topologies 
involved in a given $B$ decay channel,
therefore these corrections have to be investigated one by one.

Among the most intriguing effects in charmless $B$ decays are
the so called ``charming penguins''. 
The $c$-quark pair emitted in the $b\to c \bar{c} d (s) $ decay 
propagates in the environment of the light spectator cloud and  
annihilates to gluons, the latter being absorbed in the final 
charmless state. In this, so called BSS-mechanism \cite{BSS} 
the intermediate $c \bar{c}$ loop generates 
an imaginary part, contributing to the final-state strong rescattering phase.
In QCD factorization 
approach \cite{BBNS}, charming penguins are typically 
small, being a part of the $O(\alpha_s)$ nonfactorizable correction 
to the $B\to \pi\pi$ amplitude.  
On the other hand, fits of two-body charmless $B$ decays do not exclude
substantial $O(1/m_b)$ nonperturbative effects of the charming-penguin type
\cite{Rome}.
It is therefore important to investigate by independent 
methods the effects generated by $c$-quark loops in charmless $B$ decays.  

In this letter we report on the LCSR 
estimate of the penguin topology contributions
to the $B\to \pi\pi$ amplitude. We start with calculating 
the $c$-quark part of this effect (charming penguin).
Later on, we extend the sum rule analysis to 
the penguin contractions of the $u$-quark current-current operators. \\

{\bf 2.} As a study case we choose the $\bar{B}^0\to\pi^+\pi^-$ channel.
The decay amplitude is given by the hadronic matrix element 
$\langle \pi^+\pi^- |H_{\rm eff}|\bar{B}^0\rangle $ 
of the effective weak Hamiltonian \cite{BBL}  
\begin{eqnarray}
H_{\rm eff} &=& \frac{G_F}{\sqrt{2}} \Bigg \{ \sum_{p=u,c}
V_{pb}V^{\ast}_{pd} \left(C_1 {\cal O}_1^p + 
C_2 {\cal O}_2^p \right)- V_{tb} V^{\ast}_{td} 
\left(\sum_{i=3}^{10} C_i {\cal O}_i +C_{8g} {\cal O}_{8g}\right)\Bigg \}\,,
\label{eq:heff}
\end{eqnarray}
where  
$
{\cal O}_1^p = (\overline{d}\Gamma_{\mu}p)(\overline{p}\Gamma^{\mu} b)
$ 
and 
$
{\cal O}_2^p = (\overline{p} \Gamma_{\mu} p)(\overline{d} \Gamma^{\mu}
b)$  
are the current-current operators  
($p = u,c$ and $\Gamma_\mu=\gamma_\mu(1-\gamma_5)$),
${\cal O}_{3-10}$ are the penguin 
operators, and ${\cal O}_{8g}$ is the chromomagnetic dipole operator.
Each operator entering Eq.~(\ref{eq:heff}) contributes to the 
$B\to \pi\pi$ decay amplitude 
with a number of different contractions of the quark lines 
(topologies) \cite{BS}. 
In what follows we will mainly concentrate on the operator 
$ {\cal O}_1^c$. For convenience we decompose this operator 
$
{\cal {O}}_1^c = \frac13{\cal {O}}_2^c + 2 {\cal \tilde{O}}_2^c 
$,
extracting the colour-octet part
$
{\cal \tilde{O}}_2^c =  (\overline{c} \Gamma_{\mu} \frac{\lambda^a}{2} c)
(\overline{d} \Gamma^{\mu} \frac{\lambda^a}{2} b) \, .
$

To derive LCSR for the $B\to \pi\pi$ hadronic 
matrix element of ${\cal {O}}_1^c$
we follow the procedure which is described in detail in 
\cite{AK,KMU}. One starts from introducing the correlation function:
\begin{eqnarray}
F_{\alpha}^{(\tilde{{\cal O}}_2^c)} = 
i^2 \int \!d^4 x \,e^{-i(p-q)x}\!\int d^4 y \,e^{i(p-k)y}  \langle 0 |
T \{ j_{\alpha 5}^{(\pi)}(y) \tilde{{\cal O}}_2^c(0) j_5^{(B)}(x) \} | \pi^-(q)
\rangle 
\nonumber
\\
=(p-k)_{\alpha} F(s_1,s_2,P^2) + ...\,, 
\label{eq:corr0}
\end{eqnarray}
where 
$
j_{\alpha 5}^{(\pi)} = \overline{u} \gamma_{\alpha} \gamma_5 d
$
and
$ 
j_5^{(B)} = i m_b \overline{b} \gamma_5 d 
$ 
are  the quark currents
interpolating pion and $B$ meson, respectively. 
In the above, ellipses denote the Lorentz-structures which are not used. 
Only the colour-octet part of ${\cal O}_1^c$ needs to be taken into account. 
The contributions of ${\cal O}_2^c$ contain at least a two-gluon annihilation
of the color-neutral $c$-quark pair and have to be considered within
higher-order corrections. 
The correlation function  (\ref{eq:corr0}) depends on 
the artificial four-momentum $k$ allowing one to avoid overlaps 
of the $b$-flavoured and light-quark states in the 
dispersion relation. We also choose $p^2=k^2=0$ for simplicity and adopt
the chiral limit $q^2=m_\pi^2=0$.
Thus, the invariant amplitude $F$ is a function 
of three variables $s_1=(p-k)^2$, $s_2= (p-q)^2$ and
$P^2=(p-q-k)^2 $.

\begin{figure}
\begin{center}
\includegraphics*[width=4.5cm]{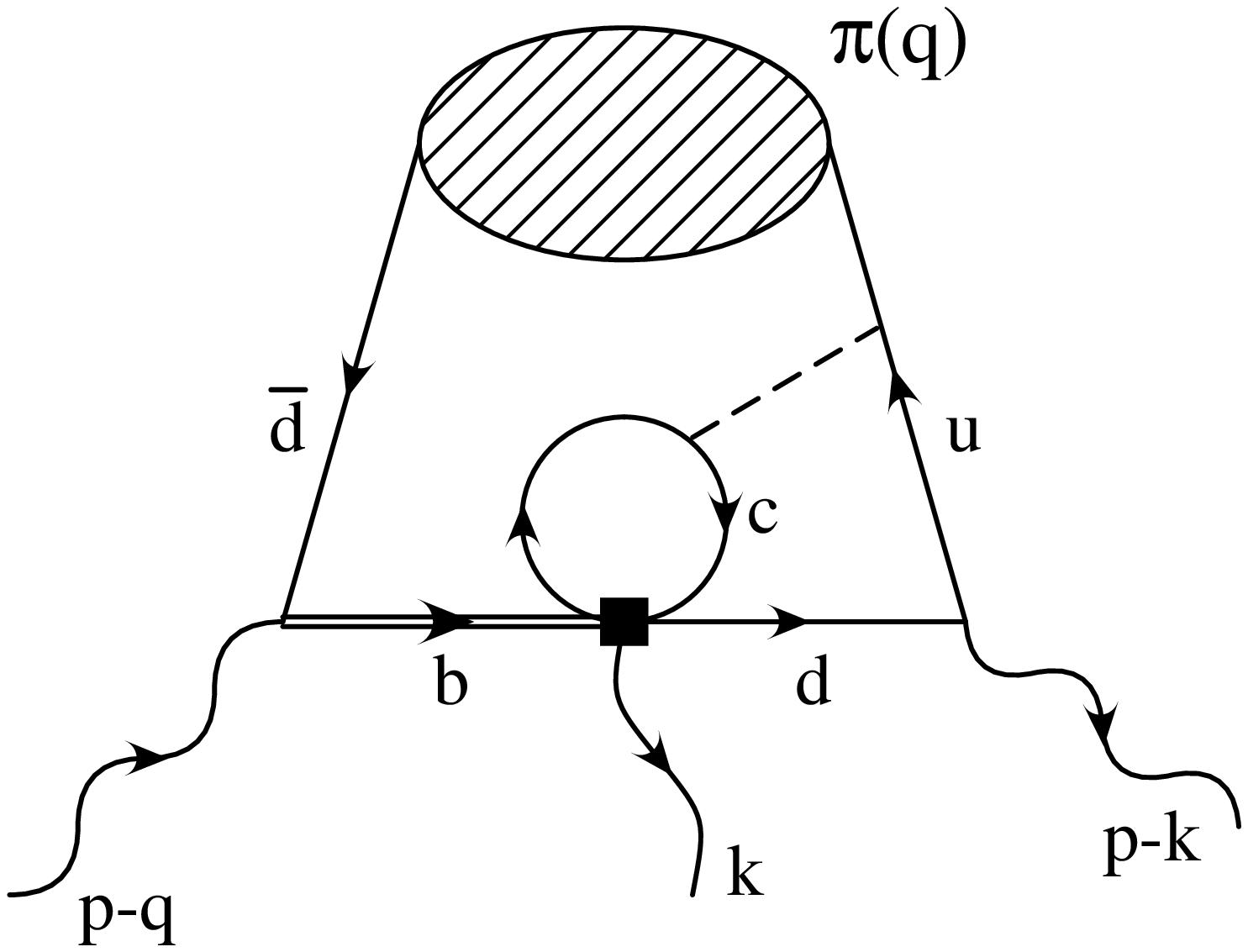} \qquad 
\includegraphics*[width=4.5cm]{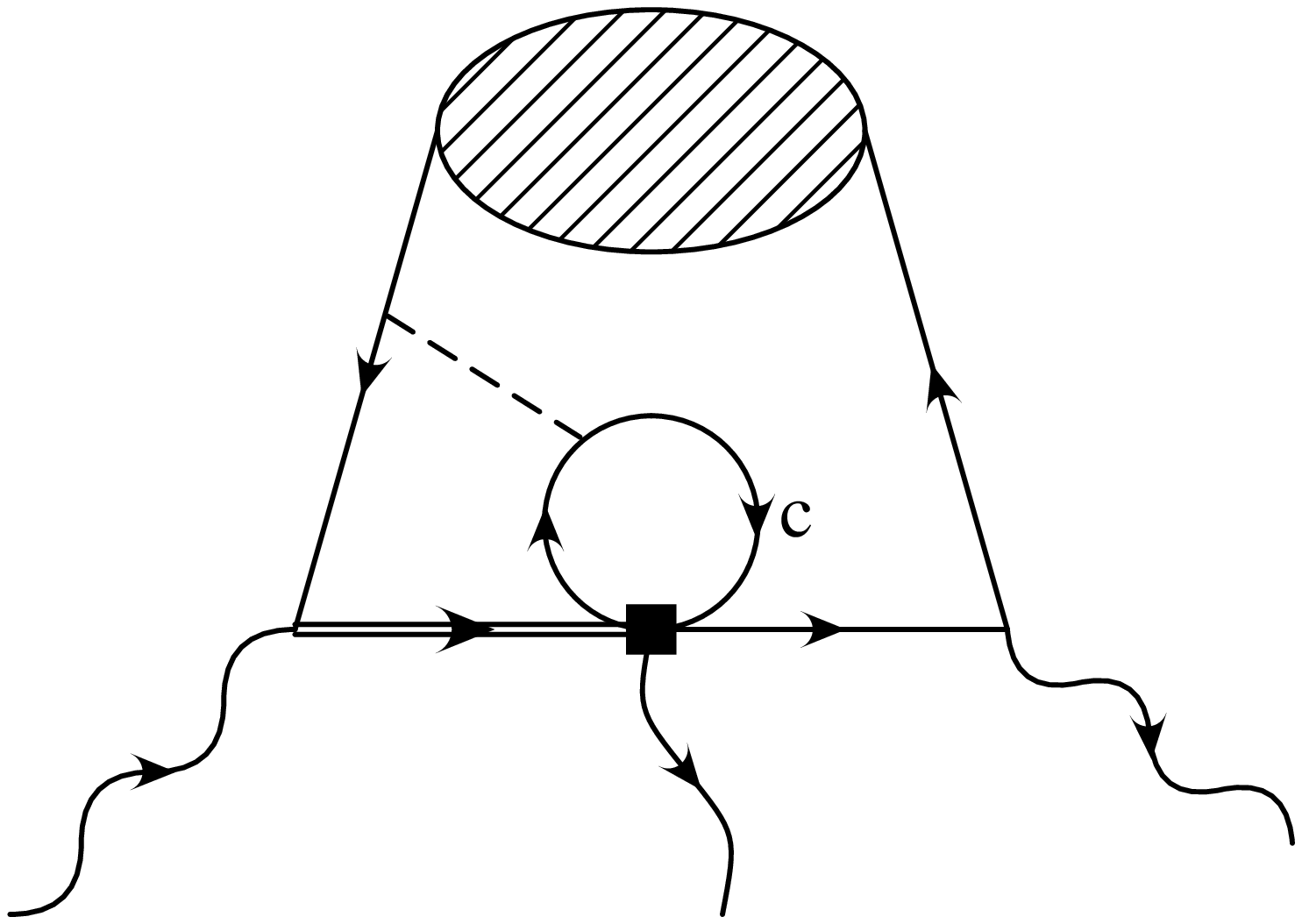}\qquad
\end{center}
\hspace{4.7cm} (a)  \hspace{5.1cm}(b) 
\caption{{\em Diagrams corresponding to 
the $O(\alpha_s)$ penguin contractions 
in the correlation function (\protect{\ref{eq:corr0}}). 
Only the diagrams contributing to the sum rule are  shown. 
The square denotes 
the four-quark operator ${\tilde{\cal O}}_2^c$. 
}}
\label{fig:hard}
\end{figure}

The next step is to calculate $F(s_1,s_2,P^2)$   at large spacelike $s_1,s_2,P^2$
employing the operator product expansion (OPE) near the light-cone.  
The corresponding diagrams of $O(\alpha_s)$  
are shown in Fig.~\ref{fig:hard}. They contain a $c$-quark loop, which
involves a well known function of $m_c^2$.
The divergence of the quark loop is absorbed in the
renormalization of the QCD penguin operators ${\cal O}_{3-6}$. 
The finite contribution of the loop is formally of the next-to-leading
order and depends on the renormalization scale $\mu$. This dependence
is compensated by the Wilson coefficients 
of the QCD penguin operators so that the remaining $\mu$ dependence is
weak, since it is of $O(\alpha_s ^2 \ln \mu)$.  
Furthermore, the finite piece of the loop contains scheme-dependent 
constants; we use the NDR scheme \cite{BBL}.

The sum rule is derived from the dispersion relations for $F$ in
the variables $s_{1}$ and $s_2$. Therefore the calculation of 
the two-loop diagrams in 
Fig.~\ref{fig:hard} 
can be considerably simplified if one starts 
from their imaginary parts obtained by employing the Cutkosky rule
and replacing the propagators by $\delta$-functions. 
For the diagram in Fig.~\ref{fig:hard}a it is sufficient
to calculate its imaginary part in $s_1$.  
Furthermore, according to the procedure
explained in \cite{AK} we have to consider the dispersion relation in
$s_1$ in the quark-hadron duality interval, 
$0\!<\!s_1\!<\!s_0^\pi\ll m_B^2$ and simultaneously at $ |P^2|\sim
m_B^2$. Hence, one  can safely neglect all $O(s_1/P^2)$
contributions in $\mbox{Im}_{s_1}F(s_1,s_2,P^2)$. In this approximation  
we obtain for the diagram in Fig.~\ref{fig:hard}a:
\begin{eqnarray}
Im_{s_1} F_{a}^{(\tilde{{\cal O}}_2^c)}(s_1,s_2,P^2) &=& 
- \frac{\alpha_s C_F f_{\pi} m_b^2}{8 \pi^2}  
 \int\limits_0^1 \frac{du}{m_b^2 - u s_2} \int\limits_0^1 dz \, I(zuP^2,m_c^2)
\nonumber \\
& & \hspace*{-2cm} 
\times P^2\Bigg \{ z(1-z) \varphi_{\pi}(u) + (1-z) \frac{\mu_{\pi}}{2 m_b}
\left [ \left (2 z + \frac{s_2}{P^2} \right )u \varphi_p(u)  
\right . 
\nonumber \\
& & \hspace*{-1.5cm} \left . + 
\left ( 2 z - \frac{s_2}{P^2}  \right ) 
\left ( \frac{\varphi_{\sigma}(u)}{3}- \frac{u\varphi_{\sigma}^{\prime}(u)}{6} \right ) 
\right ]
\Bigg \} +O\left (\frac{s_1}{P^2}\right )\,.                                  
\label{diaga}
\end{eqnarray}
In the above, $\varphi_\pi$ and $\varphi_p$, $\varphi_\sigma$ are the
pion DA of twist 2 and 
3, respectively, the latter are normalized by 
$\mu_\pi=m_\pi^2/(m_u+m_d)$
nonvanishing in the chiral limit;
we use the same standard definitions as in \cite{KMU}. Finally,  
$\varphi_\sigma^{\prime}(u)=d\varphi_\sigma(u)/du$ and $I(...)$ is the $c$-quark loop function in the NDR scheme:
\begin{eqnarray}
I(l^2,m_c^2) = \frac{1}{6} \left ( \ln \left (\frac{m_c^2}{\mu^2}\right ) + 1
\right ) + \int\limits_0^1 dx\, x (1-x) \ln \left [1 - \frac{x (1-x) l^2}{m_c^2} \right ]\,. 
\end{eqnarray} 
Similarly, for the diagram in Fig.~\ref{fig:hard}b it is sufficient 
to calculate its imaginary part in $s_2$. The result reads:
\begin{eqnarray}
Im_{s_2} F_{b}^{(\tilde{{\cal O}}_2^c)}(s_1,s_2,P^2) &=& \frac{\alpha_s C_F f_{\pi} m_b^2 }{16\pi^2}
\left (\frac{s_2 -m_b^2}{s_2} \right )^2
 \int\limits_0^1 \frac{du}{ \overline{u} P^2 + u s_1} \int\limits_0^1 dz I_c(-uz(s_2-m_b^2))
\nonumber \\
& & \hspace*{-2cm} 
\times \Bigg \{ (P^2-s_1-s_2) z \, \varphi_{\pi}(u) 
+ \frac{\mu_{\pi}}{2 m_b} (P^2-s_1+3s_2)\varphi_p(u) \nonumber \\
& & \hspace*{-1.5cm} - \frac{\mu_{\pi}}{12 m_b} \Bigg [ 
2 \left ( \frac{P^2-s_1-s_2}{u}  + 2 \frac{P^2-s_1}
{\overline{u}P^2 + u s_1}(P^2-s_1- s_2) \right ) \varphi_{\sigma}(u)  \nonumber \\
& & \hspace*{-1.5cm}  
 + (3 P^2 - 3 s_1 + s_2) \varphi_{\sigma}^{\prime}(u) \Bigg ]
\Bigg \}\,.
\label{diagb}
\end{eqnarray}  
The remaining diagrams not shown in Fig.~\ref{fig:hard},
with gluons attached to the virtual $b$ and $d$ lines,
do not contribute to the sum rule because their double imaginary
parts vanish inside the duality regions 
$0\!<\!s_1\!<s_0^\pi$, $m_b^2\!<s_2\!<s_0^B$.

Eqs.~(\ref{diaga}) and (\ref{diagb}) provide the 
leading, $O(\alpha_s)$ and 
twist 2 and 3 answer for the correlation function.
The contributions of the higher-twist pion DA's 
to $F$ are suppressed by inverse powers of  
large squares of external momenta, reflecting the power counting
in the light-cone OPE. Accordingly, in this paper, we neglect  
small effects of the twist-4 quark-antiquark DA's 
in the diagrams of Fig.~\ref{fig:hard}.

\begin{figure}
\begin{center}
\includegraphics*[width=4.5cm]{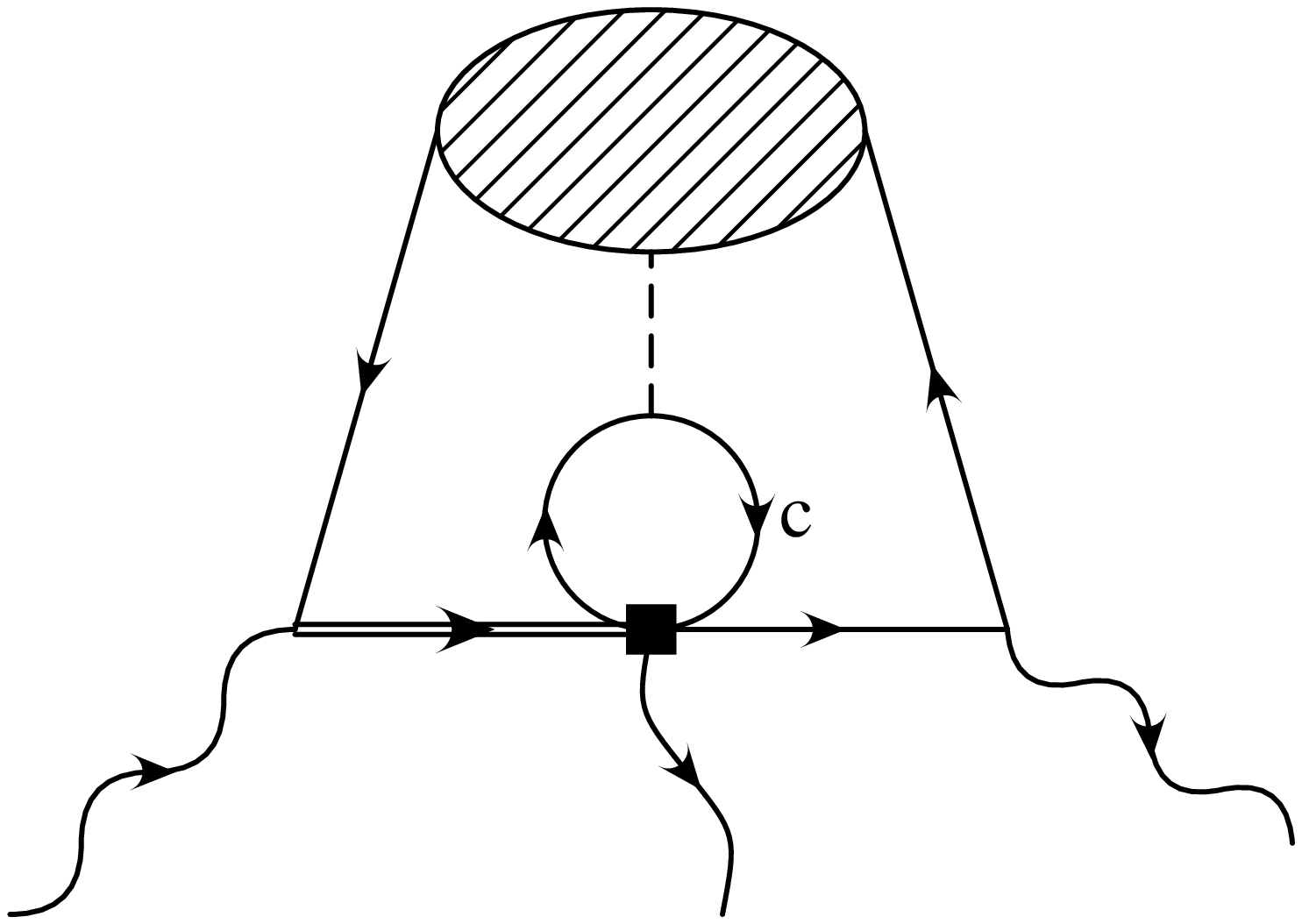}\qquad
\includegraphics*[width=4.5cm]{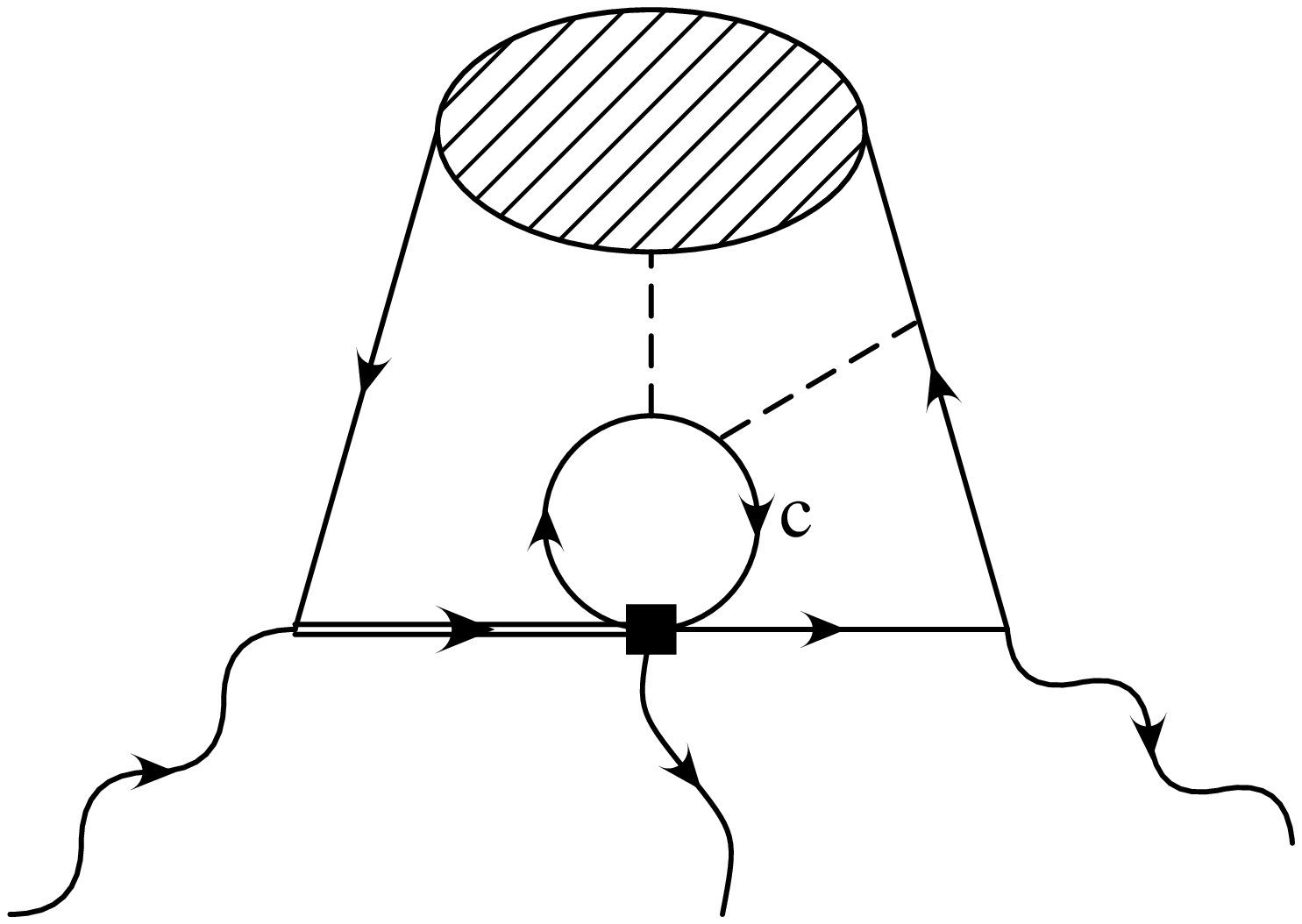} \qquad
\end{center}
\hspace{4.7cm} (a)  \hspace{5.1cm}(b) 
\caption{{\em Diagrams containing the  
quark-antiquark-gluon DA's}}
\label{fig:gluon}
\end{figure}

{\bf 3.} So far we have taken into account 
the hard-gluon emission off the c-quark loop.  
The most intriguing and difficult problem
concerning charming penguins is the effect 
of the soft (low-virtuality) gluons
coupled to the $c$-quark loop. Within the sum rule
approach this problem has to be addressed  
in the context of the light-cone OPE of the correlation function. 
Together with $\bar{u}$ and $d$ quarks, the on-shell gluons emitted at 
short distances form  multiparticle DA's of the pion. Importantly, the  contributions of these DA's 
to the sum rule, being of higher twist, are suppressed \cite{AK,KMU}
by inverse powers of the heavy mass scale
with respect to the contributions of 2-particle quark-antiquark DA's
of lower twists. Therefore, for a sum rule estimate 
of the soft-gluon effects it is sufficient to  consider diagrams
with one ``constituent'' gluon i.e., diagrams involving  
quark-antiquark-gluon DA's of the pion.  For the gluonic
penguin operator, this soft-gluon effect turns out to be  important \cite{KMU},
because in the quark-antiquark-gluon term of the sum rule 
the $1/m_b^2 $ suppression is compensated by the absence of $\alpha_s$.
In the correlation function (\ref{eq:corr0}) the situation is 
different. The diagram with one gluon shown in Fig~\ref{fig:gluon}a 
vanishes due to the current conservation in the $c$-quark loop.
Nonvanishing terms with the three-particle DA's emerge from the diagrams, 
containing at least one hard gluon in addition to  the on-shell gluon. One
of these diagrams  is shown in Fig~\ref{fig:gluon}b. 
Their complete calculation is a difficult task. From the studies of 
the $b\to s\gamma$ matrix elements of ${\cal O}_{1,2}$, 
where similar diagrams with an
on-shell photon and virtual gluon have been calculated \cite{bsgamma}, 
we conclude that the contribution of Fig.~\ref{fig:gluon}b and similar diagrams  
not only contain $\alpha_s$ but also have an 
additional $O(1/m_b^2)$  suppression 
with respect to the diagrams in Fig.~\ref{fig:hard}.

\begin{figure}
\begin{center}
\includegraphics*[width=4.5cm]{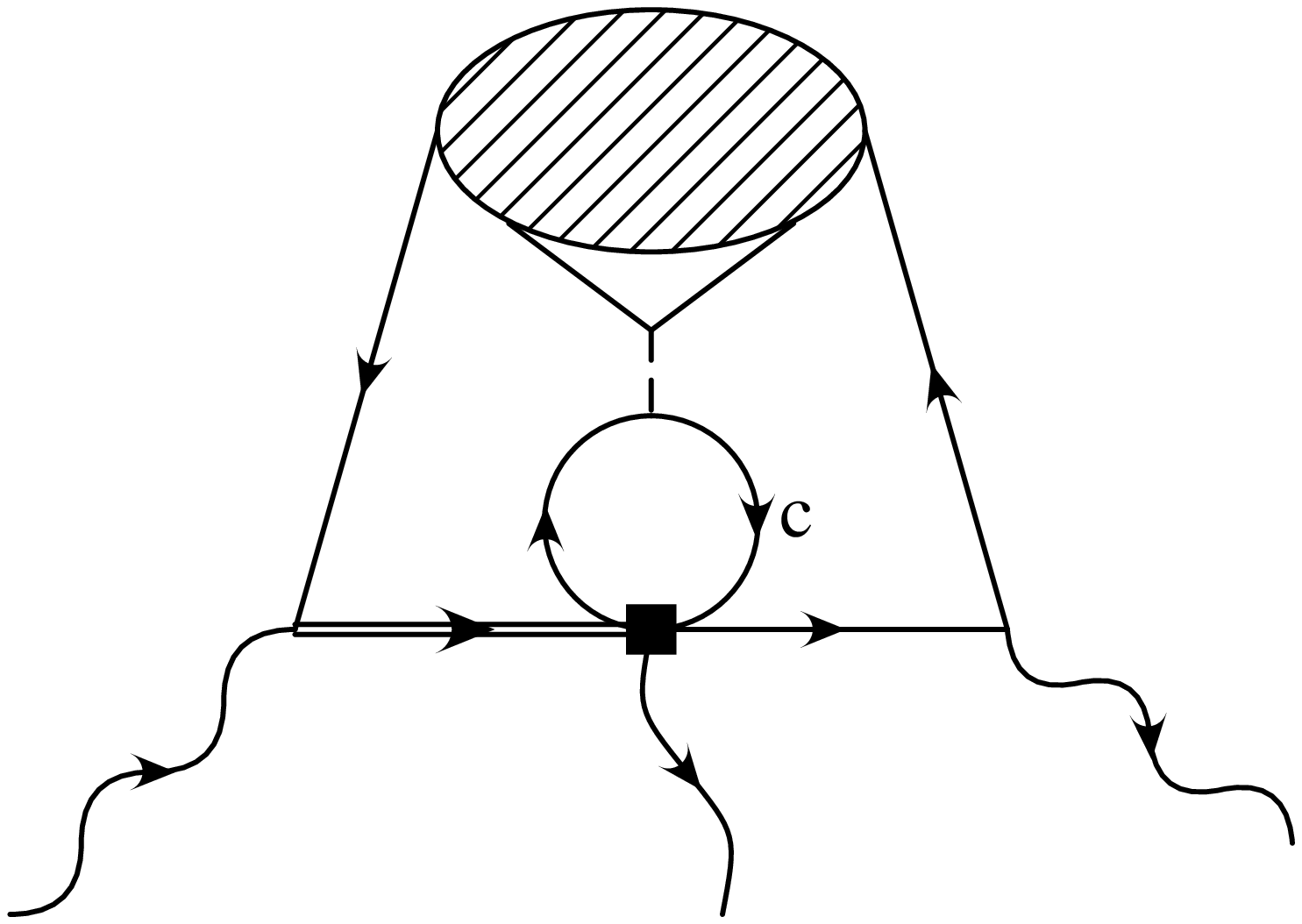}\qquad
\includegraphics*[width=4.5cm]{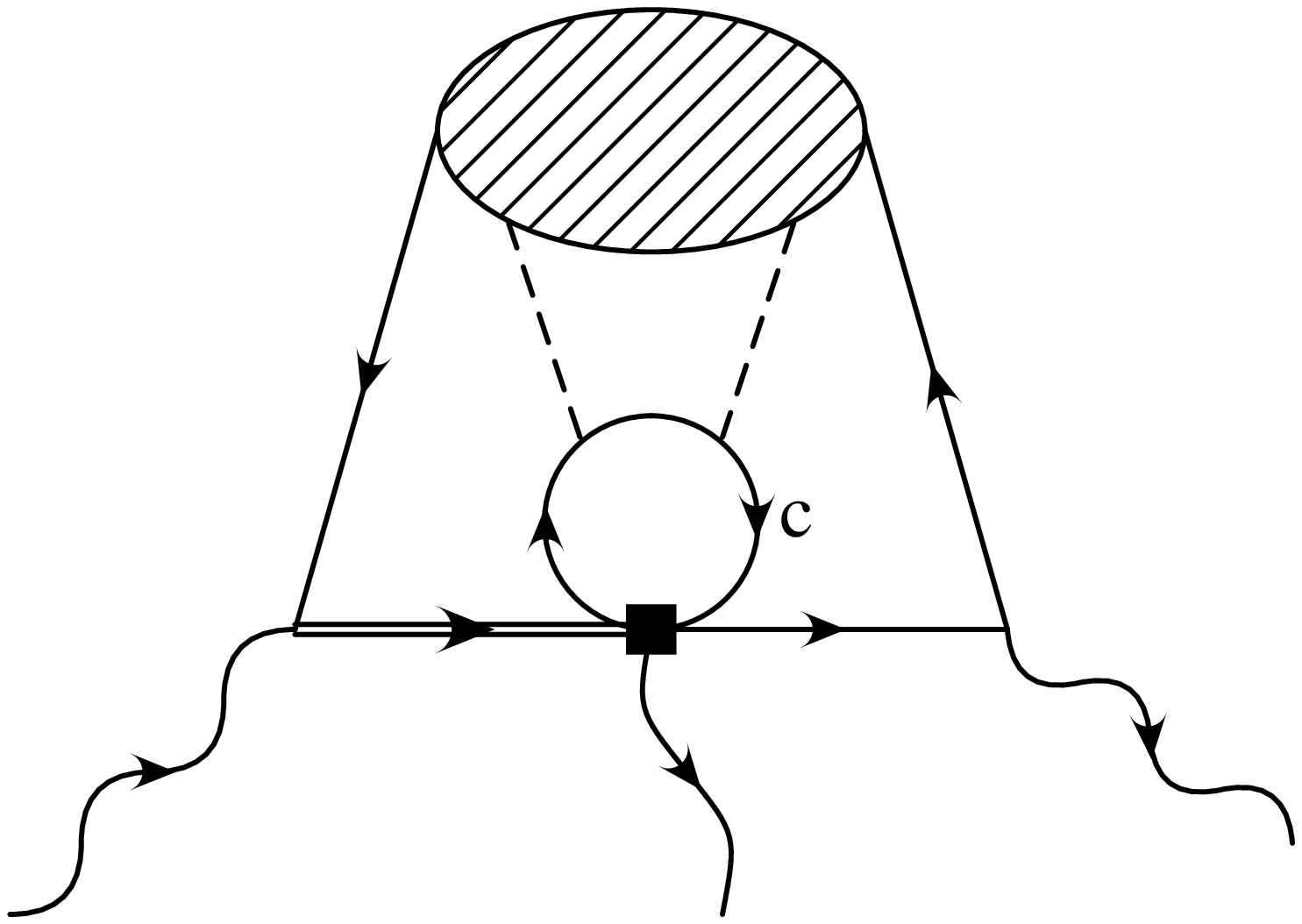} \qquad
\includegraphics*[width=4.5cm]{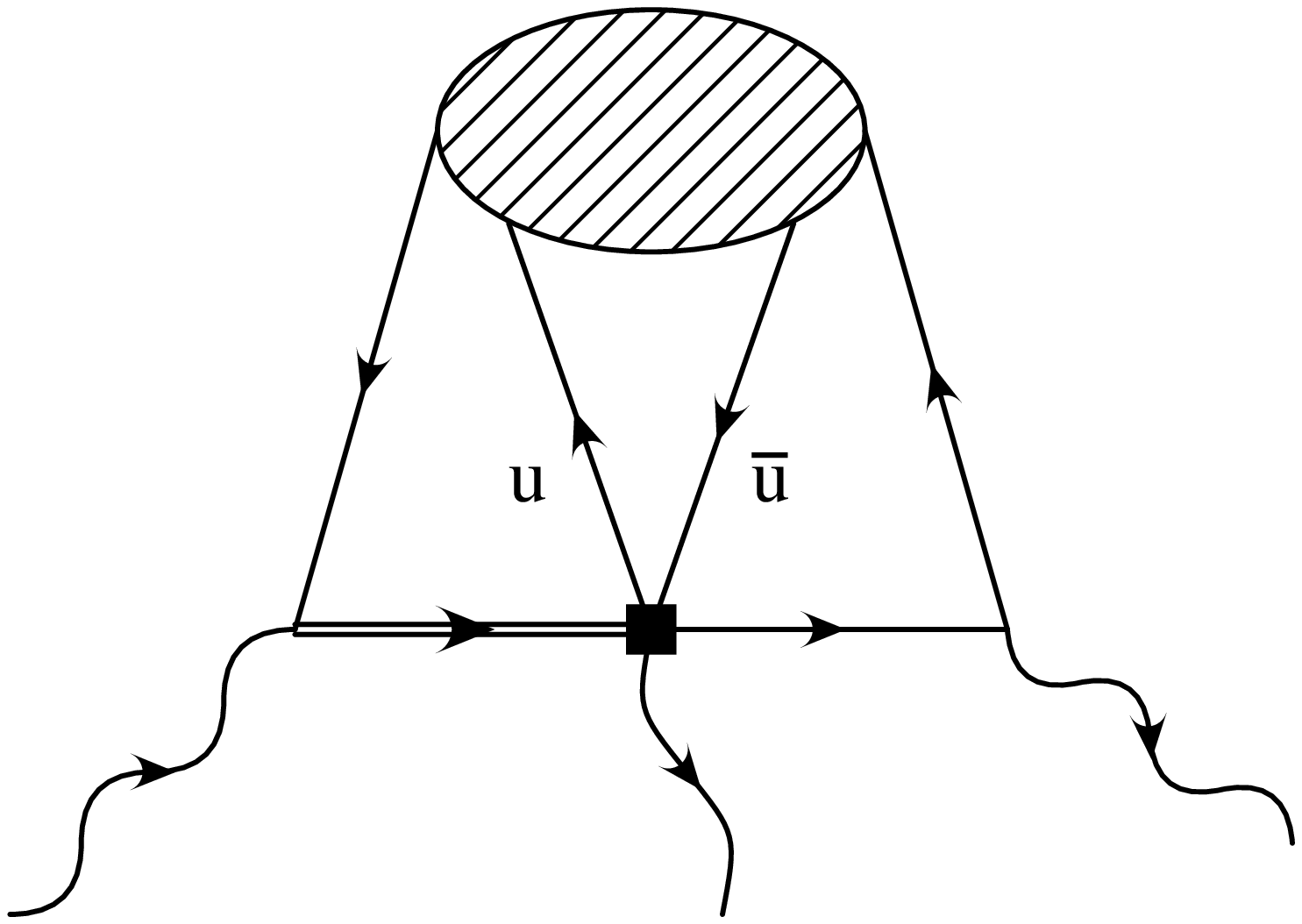} \qquad
\end{center}
\hspace{2.3cm} (a)  \hspace{4.5cm}(b)  \hspace{5.2cm}(c) 
\caption{{\em (a),(b) Diagrams corresponding to the multiparticle DA's
of the pion and arising from the expansion of the $c$-quark loop in the external gluon
field; (c) the four-quark ``soft'' contribution in the case of the
operators $O_{1,2}^u$.} }
\label{fig:4q}
\end{figure}

The next nonvanishing contribution  
to the expansion of the coloured $c$-quark loop near the light-cone
contains a  derivative of the gluon
field $D_\nu G^a_{\mu\nu}$ which can be further reduced to the light-quark pair
due to QCD equation of motion. 
The whole effect has a short-distance nature, similar to 
the formation of the QCD penguin operators.
The resulting diagram shown in Fig.~\ref{fig:4q}a
has to be calculated in terms 
of the pion four-quark DA's and is beyond the approximation adopted here
and in \cite{KMU}. Simple dimension counting yields 
for this diagram a suppression factor of $O(1/m_b^3)$. 
The $c$-loop factor in this case reduces to  
$\ln(m_c^2/\mu^2)$ plus a scheme-dependent constant. The $\mu$- and
scheme-dependence
are compensated by the contributions of QCD penguin operators,
since the c-quark loop momenta larger than $\mu$ are included in 
the short-distance coefficients of these operators. 

The diagram with two gluons emitted from the $c$-quark loop
(Fig.~\ref{fig:4q}b) 
is also not included in our calculation because it contains DA's 
with multiplicity larger than three. Here both color-neutral and color-octet parts
of ${\cal O}_1$ (that is, both ${\cal O}_2$ and $\tilde{{\cal O}}_2$) 
contribute. For simplicity we concentrate
on ${\cal O}_2$ where only the axial-vector part of the $c$-quark
current is relevant. 
Since all light degrees of freedom 
in this diagram enter DA of the massless pion ($q^2=0$), it is clear that 
the four-momentum flowing through the $c$-quark current 
has to be lightlike, i.e. the $c$-loop remains 
far off-shell (with a typical
virtuality scale of $2m_c$ independent of $m_b$).
Hence, one can use the local expansion of the c-quark 
axial-vector current in the gluon field obtained in \cite{intrinsic}:
%\begin{eqnarray}
$\bar{c}\gamma_\mu \gamma_5c
\rightarrow g_s^2/(192\pi^2 m_c^2)
\left[\partial_\mu (G_{\rho\nu}^a
\widetilde{G}^{a \rho\nu})-4(D_\alpha G^{a,\nu\alpha})
\widetilde{G}^a_{\mu\nu} \right]
$.
%\end{eqnarray}
The suppression factor with respect to the diagrams 
in Fig.~\ref{fig:hard} obtained by counting dimensions
of the resulting operators (DA's) 
is for this contribution at least $O(1/(m_c^2m_b^2))$. The presence of 
$\ln(m_c)$ and $m_c^{-2}$ in the contributions of Fig.\ref{fig:4q}a and
3b respectively, indicates that at $m_c\to 0$  
these terms are divergent. In other words, if $c$ quarks
are replaced with the light quarks, e.g., in the case 
of ${\cal O}_1^u$, the light-quark pair propagates 
at long distances, that is, belongs to the pion four-quark DA. The
corresponding diagram is the one shown in Fig.~\ref{fig:4q}c.

\begin{figure}
\begin{center}
\includegraphics*[width=4.5cm]{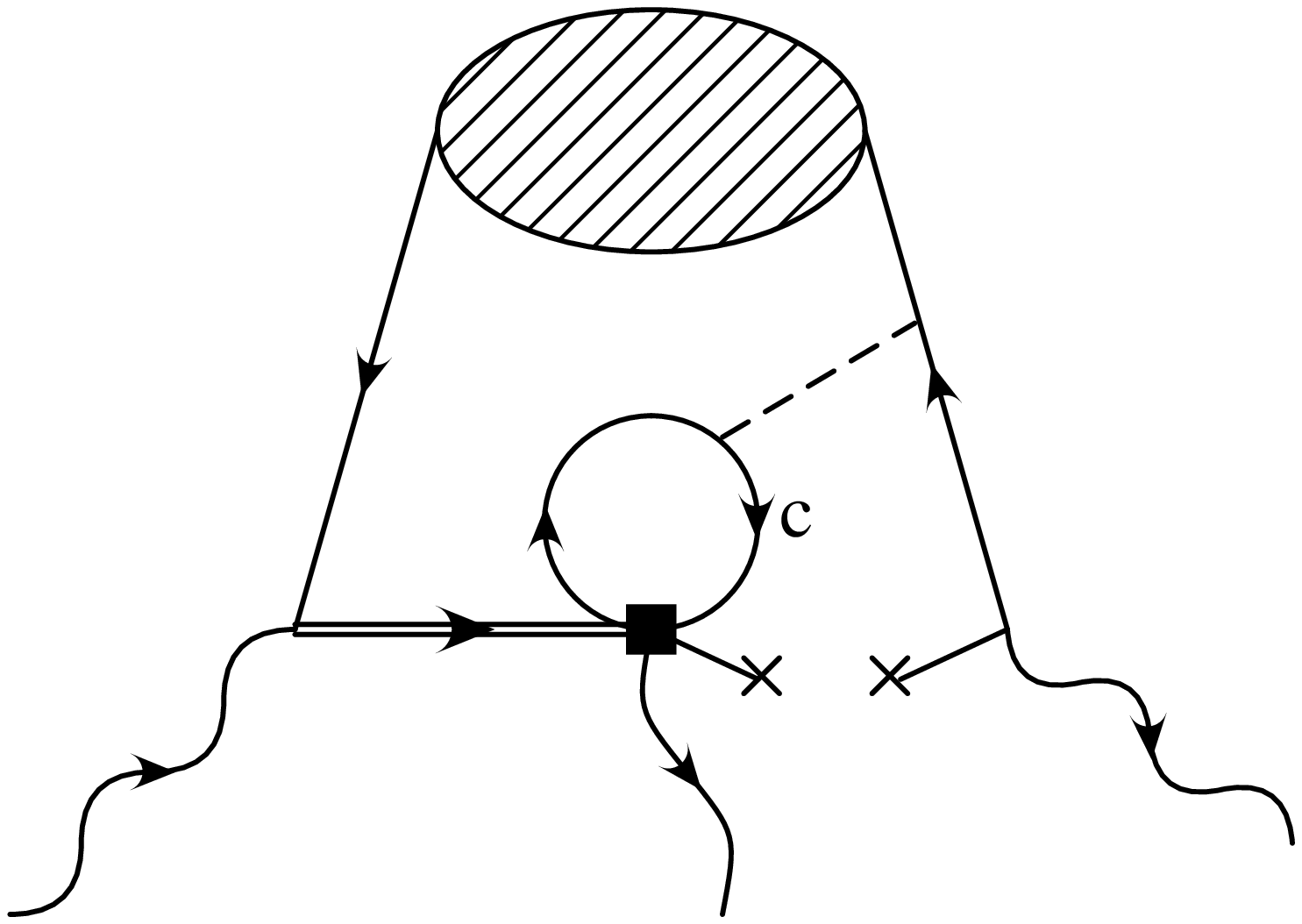} \qquad 
\includegraphics*[width=4.5cm]{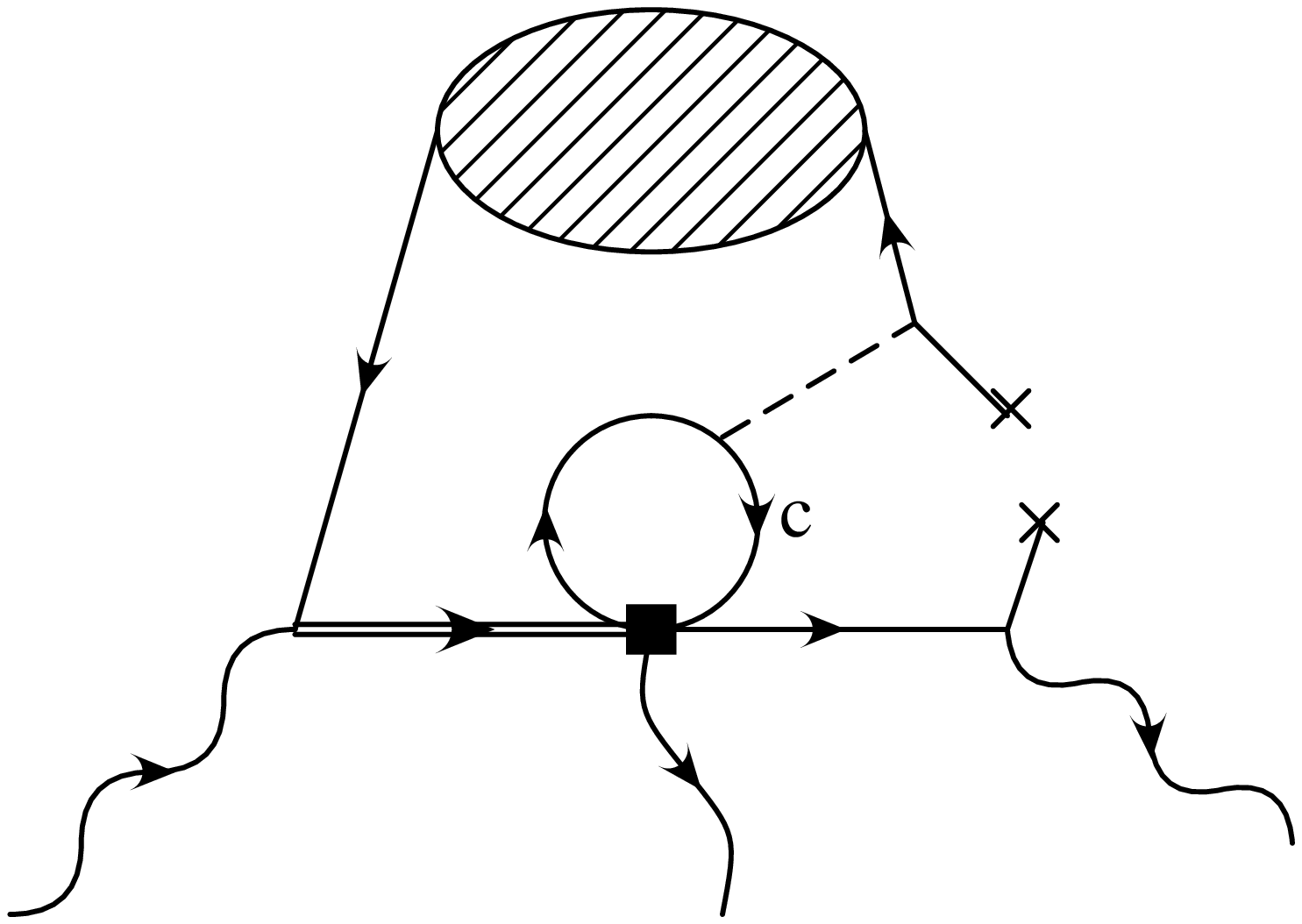}
\end{center}
\hspace{4.7cm} (a)  \hspace{5.1cm}(b) 
\caption{{\em Diagram corresponding to the factorizable 4-quark contribution 
to the correlation function.}}
\label{fig:cond}
\end{figure}

As already mentioned, we neglect the contributions of  
four-quark DA's stemming from the matrix elements of the type 
$\langle 0 \mid \bar{u}(x_1)\bar{q}(x_2)q(x_3) d(x_4) |\pi\rangle$ 
($x_i$ on the light-cone).  On the other hand, following \cite{KMU}
we take into account the factorizable parts of the 4-quark
vacuum-pion matrix elements, extracting the configurations 
where one quark-antiquark pair forms the quark vacuum condensate, whereas the other one 
hadronizes into a twist 2 and 3 pion DA. Such 
contributions are enhanced by the  large parameter $\mu_\pi$. 
In the approximation adopted here, only  
two diagrams shown in Fig.~\ref{fig:cond} contribute to the sum
rule (that is, have a nonvanishing contribution to the double dispersion
relation in the duality region). Note that  the quark-condensate diagram
in Fig.~\ref{fig:cond}b 
originates from the 4-quark diagram in
Fig.~\ref{fig:4q}a. The diagrams in Fig.~\ref{fig:cond} have only 
one loop and their calculation is relatively simple yielding the
following result:
\begin{eqnarray}
F_{\langle q \overline{q} \rangle}^{(\tilde{{\cal O}}_2^c)}(s_1,s_2,P^2) = 
- \frac{\alpha_s C_F f_{\pi} m_b \langle q \overline{q} \rangle}{12 \pi} 
 \int\limits_0^1 \frac{du}{(m_b^2 - u s_2) s_1} \Bigg\{I(u P^2 + \overline{u} s_1,m_c^2)
\Bigg [ 
2 s_2 \varphi_{\pi}(u) 
\nonumber \\
+ 3 \mu_{\pi} m_b \varphi_{p}(u) 
\!+\!\frac{\mu_{\pi}}{6 m_b} 
\Bigg ( \left [
\frac{P^2-s_1}{u P^2 + \overline{u} s_1} \!\left (\! 2 + \frac{u s_2}{m_b^2 - u s_2} \right )\! - 
 \frac{s_2}{m_b^2 - u s_2} \right ]\varphi_{\sigma}(u)\!  - \varphi_{\sigma}^{\prime}(u)\!
\Bigg )\! \!\Bigg ]
\nonumber \\
-I(0,m_c^2)\Bigg[ \frac{P^2-3s_2-s_1}2\varphi_\pi(u)
-\mu_\pi m_b\left(3\varphi_p(u)+\frac{P^2-s_1}{6s_2}
\varphi_{\sigma}^{\prime}(u)\right)\Bigg]\Bigg\}.
\label{eq:cond}
\end{eqnarray}
Multiparticle contributions which are factorized in the condensates of higher dimension 
are not taken into account. We also neglect the quark-condensate
contributions of the type $\langle \bar{q}q\rangle \langle 0\mid
\bar{u}(x_1)G_{\mu\nu}^a(x_2) d(x_3)\mid \pi \rangle$ arising from 
the diagram in Fig.~\ref{fig:4q}b after applying the QCD equation of
motion to the derivatives of $G_{\mu\nu}^a$. These terms are suppressed 
at least by $O(1/m_c^2)$ with respect to the diagrams in Fig.~\ref{fig:cond}.
 
Summarizing, we do not find significant contributions involving soft
gluons in the OPE of the correlation function (\ref{eq:corr0}). 
The dominant effect arises from the $c$-quark loop annihilation into 
hard gluons (Fig.~\ref{fig:hard}). In addition, there is the
quark-condensate contribution (Fig.~\ref{fig:cond}) which we consider 
a natural upper limit for all neglected contributions of multiparticle DA's. 
Importantly, the presence of the new intermediate scale $m_c\ll m_b$ 
in the correlation function does not noticeably enhance the charming penguin
contribution: 
in the leading terms of OPE $m_c$ enters logarithmically, and  
the inverse powers of $m_c$ appear only in the subleading suppressed terms.

{\bf 4.} Having calculated the relevant contributions to the correlation
function we are now in a position to obtain 
the sum rule for the hadronic matrix element 
$
A^{(\tilde{{\cal O}}_2^c)}( \bar{B}^0_d \to \pi^+ \pi^-) 
\equiv \langle \pi^-(p)\pi^+(-q) |
\tilde{{\cal O}}_2^c |\bar{B}^0_d(p-q) \rangle
$ . 
The derivation of LCSR explained in detail in \cite{AK,KMU}
includes: use of the dispersion relations for the invariant amplitude 
$F(s_1,s_2,P^2)$ in the variables $s_1$ and $s_2\,$; 
employing quark-hadron duality in both pion and $B$ meson channels
with the threshold parameters $s_0^\pi$ and $s_0^B$, respectively; 
the Borel transformations and 
the continuation  from large spacelike $P^2$ to large timelike
$P^2=m_B^2$. As a result, LCSR is given by the following expression: 
\begin{eqnarray}
&&f_\pi f_B A^{(\tilde{{\cal O}}_2^c)}( \bar{B}^0_d \to \pi^+ \pi^-)
e^{-m_B^2/M_2^2}  
\nonumber\\
&&=\int\limits_{m_b^2}^{s_0^B} \!ds_2
e^{-s_2/M_2^2}\!
\Bigg \{
\!\int\limits_0^{s_0^\pi} ds_1 e^{-s_1/M_1^2}\mbox{Im}_{s_2}\mbox{Im}_{s_1} F(s_1,s_2,P^2)
\Bigg\}_{P^2\to m_B^2}\,,
\label{eq:sumrule}
\end{eqnarray}
where $M_{1}$ and $M_2$ are the Borel parameters in the pion and
$B$-meson channels, respectively.  
Using Eqs.~(\ref{diaga}),(\ref{diagb}) and (\ref{eq:cond}),
we obtain from Eq.~(\ref{eq:sumrule}) the ``charming penguin''
hadronic matrix element at $O(\alpha_s)$ and with twist 2 and 3 accuracy :
\begin{eqnarray}
A^{(\tilde{{\cal O}}_2^c)}( \bar{B}^0_d \to \pi^+ \pi^-)
=A_{Fig.1}+A_{\langle\bar{q}q\rangle}\,,
\label{eq:srfinal}
\end{eqnarray}
\\
where 
\begin{eqnarray}
A_{Fig.1}&=&
i\frac{\alpha_s C_F}{\pi} m_B^2 \left(\int\limits_0^{s_0^\pi}
\frac{ds e^{-s/M_1^2}}{4\pi^2f_\pi}\right)
\frac{m_b^2 f_{\pi}}{2 m_B^2f_B}\int\limits_{u_0^B}^1 \frac{du}{u} 
e^{(m_B^2 - \frac{m_b^2}{u})/M_2^2 }
\nonumber \\
& & \hspace*{-1.8cm} 
\times \Bigg \{ 
\int\limits_0^1 dz \, I(z m_B^2u,m_c^2)
\Bigg [ z(1-z) \varphi_{\pi}(u) 
\nonumber \\
& & \hspace*{-1.3cm} +(1-z) \frac{\mu_{\pi}}{2 m_b} \left [ 
u \varphi_p(u) \left ( 2 z + \frac{m_b^2}{u m_B^2} \right ) 
+ 
\left ( \frac{\varphi_{\sigma}(u)}{3}- \frac{u\varphi_{\sigma}^{\prime}(u)}{6}
\right ) \left ( 2 z - \frac{m_b^2}{u m_B^2}  \right ) \right ]
\Bigg ] 
\nonumber \\
& & \hspace*{-1.4cm} -  
\frac{m_b \mu_{\pi}}{4 m_B^2} \int\limits_0^1 dz \, I(-z m_b^2 \overline{u}/u)\frac{\overline{u}^2}{u}
\Bigg [\varphi_p(1) \left (1 + \frac{3m_b^2}{u m_B^2} \right ) + 
\frac{\varphi_{\sigma}^{\prime}(1)}{6} \left (1 - \frac{5m_b^2}{u m_B^2} \right )
\Bigg ] 
\Bigg \}
\label{eq:fig1sr}
\end{eqnarray}
is the contribution of the diagrams in Fig.~\ref{fig:hard}
and 
\begin{eqnarray}
A_{\langle q \overline{q} \rangle} &=&
i\frac{\alpha_s C_F}{6 \pi} m_b^2 
\left ( - \frac{\langle q \overline{q} \rangle}{f_{\pi} m_b} \right ) 
\left ( \frac{m_b^2 f_{\pi}}{2 m_B^2 f_B} \right ) 
\int\limits_{u_0^B}^1 \frac{du}{u^2} e^{(m_B^2 - \frac{m_b^2}{u})/M_2^2
}\Bigg\{ I(u m_B^2,m_c^2)
\nonumber \\
& & \hspace*{-1cm}
\times \Bigg [
2 \varphi_{\pi}(u) + 
\frac{\mu_{\pi}}{m_b} \left( 3u \varphi_{p}(u) 
+ \frac{\varphi_{\sigma}(u)}{3} - \frac{u \varphi_{\sigma}^{\prime}(u)}{6} 
\right)  
\Bigg ]
\nonumber \\
& & \hspace*{-1cm}
+I(0,m_c^2)\Bigg[ \frac{\varphi_\pi}2\left(
  3-\frac{m_B^2u}{m_b^2}\right)+
\frac{\mu_\pi}{m_b}\left(3u\varphi_p(u)+
\frac{m_B^2u^2}{6m_b^2}\varphi_{\sigma}^{\prime}(u)\right)\Bigg]\Bigg\}
\label{eq:qqsr}
\end{eqnarray}           
is the factorizable 4-quark contribution.  In the above, $u_0^B=m_b^2/s_0^B$.
The sum rule (\ref{eq:qqsr}) is obtained at finite $m_b$ 
but we neglect numerically very small corrections of order $s_0^\pi/m_B^2$.
Note that the $c$-quark loop factor $I(zm_B^2u,m_c^2)$ in the sum rule
originating from the Fig.~\ref{fig:hard}a  diagram  
has a complex phase which is generated by the analytic continuation
in $P^2$ and has to be associated with the strong-interaction phase 
(in the quark-hadron duality approximation). The BSS mechanism \cite{BSS} 
is thus recovered in LCSR.

To compare our result with  
QCD factorization we investigate the heavy-quark mass 
limit of the sum rule (\ref{eq:srfinal}),
expanding all heavy-mass dependent quantities
in powers of $m_b$. At $m_b\to \infty $  
only the diagram of Fig.~\ref{fig:hard}a contributes. In this limit
duality interval of the $u$-integration in Eq.~(\ref{eq:fig1sr}) reduces to
the end-point region allowing one to put
$u\to 1$ in the $c$-quark loop factor. After that, the 
integral over $u$ multiplying this factor 
reproduces LCSR for the $B\to \pi$ form factor \cite{Bpi}. To see that 
one has to use the relation \cite{BF} following from the QCD equation of
motion : 
$u\varphi_p(u)+\varphi_\sigma(u)/3-u\varphi_\sigma^{\prime}(u)/6=0$ 
(neglecting the quark-antiquark gluon DA's). After these transformations 
we obtain :
\begin{equation}
\hspace*{-0.2cm}
A^{(\tilde{{\cal O}}_2^c)}( \bar{B}^0_d \to \pi^+ \pi^-)
\Bigg|_{m_b\to \infty}\!\!\!\!= \frac{\alpha_s C_F}{\pi}
\Bigg(\int\limits_0^1dz
z(1-z) I(zm_b^2,m_c^2)\Bigg)\Big(i \sqrt{m_b}f_\pi^{(SVZ)} \hat{f}_{B\pi}^+(0)^{(LCSR)}\Big)\,.
\label{eq:hql}
\end{equation}
In the above, we have denoted the integral over $s$ in Eq.~(\ref{eq:fig1sr})
by $f_\pi^{(SVZ)}$ indicating that it coincides with the leading-order 
SVZ sum rule for $f_\pi$ \cite{SVZ}. Furthermore we have replaced the
LCSR $B\to\pi$ form factor by an $m_b$-independent effective form factor  
$f_{B\pi}^+(0)^{(LCSR)}=\hat{f}_{B\pi}^+(0)^{(LCSR)}/m_b^{3/2}$.
Thus, the charming penguin matrix element at $m_b\to \infty$
factorizes into the c-loop integral and the factorizable 
$B\to \pi\pi$ amplitude equal to the matrix element of ${\cal O}_1^u$ in the emission topology:
$A^{({\cal O}_1^u)}_E( \bar{B}^0_d \to
\pi^+ \pi^-)_{m_b\to \infty}=  i\sqrt{m_b}f_\pi \hat{f}_{B\pi}^+(0)$.
Furthermore, since the asymptotic twist-2 pion DA has the form 
$\varphi_\pi(z)=6z(1-z)$, the limiting expression (\ref{eq:hql})
reproduces the c-quark penguin-loop contribution 
to the $B\to \pi\pi$ amplitude in the QCD factorization \cite{BBNS} at
the twist 2 level.  
For brevity, we omit a more detailed discussion of various $1/m_b$ terms
which at the end lead to a numerical deviation of the LCSR (\ref{eq:srfinal})
from the $m_b\to \infty $ limit. 
We only note that the contribution of the diagram in
Fig.~\ref{fig:hard}b which does not have a counterpart in QCD
factorization is suppressed as $1/m_b^2$. Strictly speaking this
diagram turns out to be beyond the adopted accuracy 
since  in LCSR we have neglected other small $O(1/m_b^2)$ terms.

We conclude this section with a comment on the penguin contractions 
of the current-current operator ${\cal O}_{1}^u$. 
The corresponding sum rule for the  
hadronic matrix element  
$
A^{(\tilde{{\cal O}}_2^u)}_P( \bar{B}^0_d \to \pi^+ \pi^-) 
\equiv \langle \pi^-\pi^+|
\tilde{{\cal O}}_2^u |\bar{B}^0_d \rangle_P
$ is easily obtained from LCSR ~(\ref{eq:srfinal})
by putting $m_c\to 0$ everywhere except in the second term of the 
quark condensate contribution (\ref{eq:qqsr}). As we already discussed
this term emerges from the factorization of the diagram in Fig.~\ref{fig:4q}a
and in the case of the light quarks it has to be absorbed in the
four-quark DA's (Fig.~\ref{fig:4q}c).  This four-quark DA contribution does not contain 
any $\alpha_s$ suppression. However, counting the dimensions we find that 
it is power-suppressed, at least by $O(1/m_b^3)$.

{\bf 5.} For a numerical estimate of the charming
penguin in $B\to \pi\pi$ decay we calculate 
the ratio of the sum rule (\ref{eq:srfinal})
to the factorizable amplitude 
$A^{({\cal O}_1^u)}_E( \bar{B}^0_d \to \pi^+ \pi^-) 
=im_B^2f_\pi f^+_{B\pi}(0)$:
\begin{equation}
r^{({\cal O}_1^c)}(\bar{B}^0_d \to \pi^+ \pi^-)\equiv 
\frac{A^{({\cal O}_1^c)}( \bar{B}^0_d \to \pi^+ \pi^-)}{
A^{(O_1^u)}_E( \bar{B}^0_d \to \pi^+ \pi^-)}
\simeq 
\frac{2A^{(\tilde{{\cal O}}_2^c)}( \bar{B}^0_d \to \pi^+ \pi^-)}{
im_B^2f_\pi f^+_{B\pi}(0)^{(LCSR)}}\;.
\label{eq:ratio}
\end{equation}
We use the same input as in \cite{KMU} and in the numerical
analysis of the LCSR for the $B\to \pi$ form factor \cite{Bpi}. 
The following parameters are taken for the pion channel: 
$f_\pi=132$ MeV, $s_0^\pi=0.7$ GeV$^2$,
$M^2_1= 0.5\,\mbox{-}1.5 $ GeV$^2$; and for the $B$ channel:
$m_b=4.7 \pm 0.1$ GeV (the one-loop pole mass), $s_0^B= 35\mp 2$ GeV$^2$,   
$M_2^2=8\,\mbox{-} 12$ GeV$^2$. 
The normalization scale adopted for the pion DA's and $\alpha_s$ is 
$\mu_b= \sqrt{m_B^2-m_b^2}\simeq 2.4$ GeV.
For the  $c$ quark mass we take $m_c=1.3\pm 0.1$ GeV, and for the quark condensate density   
$\langle \bar{q}q \rangle(1 \mbox{GeV})=-(240\pm 10 ~\mbox{MeV})^3$,
or, equivalently, $\mu_\pi(1 \mbox{GeV})= 1.59 \pm 0.2 $ GeV. 
All pion light-cone DA's are taken in the asymptotic form:
$\varphi_\pi(u)=\varphi_\sigma(u)=6u(1-u),~\varphi_p(u)=1$.
To get a feeling how the nonasymptotic form of the pion DA 
influences our result we also recalculate Eq.~(\ref{eq:ratio})
using a model for $\varphi_\pi(u)$ with  a nonzero second Gegenbauer
coefficient $a_2(1 \mbox{GeV})=0.4$.  With the above input 
and adding the uncertainties caused by the variation of all parameters  linearly, 
we get the following range:
\begin{equation}
r^{({\cal O}_1^c)}(\bar{B}^0_d\to \pi^+\pi^-)
= [-(0.29 \div 0.56) -(1.3 \div 1.6)i]\cdot 10^{-2}\,.
\label{numberc}
\end{equation}
The corresponding estimate for the penguin contraction 
of the operator ${\cal O}_1^u$ is: 
\begin{equation}
r_P^{({\cal O}_1^u)}(\bar{B}^0_d\to \pi^+\pi^-)= 
[(0.09\div 0.21) -(1.6\pm 2.1)i]\cdot 10^{-2}\,.
\label{numberu}
\end{equation}
  
The penguin-topology contributions turn out to be very small,
not larger than the other nonfactorizable corrections in $B\to \pi\pi$, 
however, as  we shall see in the next section, they add up  
to a noticeable effect in the CP asymmetry. The 
contribution of the quark condensate term in the sum rule 
to the real parts of Eqs.~(\ref{numberc}) and (\ref{numberu})
is relatively small,  but constitutes about 50\% of
the imaginary parts. 

{\bf 6.} It is well known that the penguin
contributions play a key role in the direct and mixing-induced CP violation in $B \to \pi \pi$. 
The time-dependent CP-asymmetry is given by 
\begin{eqnarray}
{\cal A}_{\rm CP}(B^0_d\to \pi^+\pi^-) (t) 
 = a_{\rm CP}^{\rm dir} \cos(\Delta M_d\,t)+ a_{\rm CP}^{\rm mix}
\sin(\Delta M_d\,t)\label{e1}\,,
\end{eqnarray}
where $a_{\rm CP}^{\rm dir} 
\equiv (1-\left\vert \xi \right\vert^2)/(1+\left\vert \xi \right\vert^2)$
and 
$a_{\rm CP}^{\rm mix} \equiv (2\,\mbox{Im}\,\xi)/(1+\left\vert\xi\right\vert^2)$,
with $\xi = e^{-2i(\beta+\gamma)}(1 + R\,  e^{i\gamma})/(1 + R \, e^{-i\gamma})$
and $R \equiv -P/(R_bT)$.
Here $T$ is the contribution to the $B\to \pi\pi$ amplitude proportional to
$V_{ub} V_{ud}^* = |V_{ub} V_{ud}^*| e^{-i\gamma} $ and contains also
the penguin contraction of the current-current
operator ${\cal O}_1^u$. The remaining amplitude $P$ is
proportional to $V_{cb} V_{cd}^*$. 
The factor $R_b = |V_{ub}||V_{ud}|/(|V_{cb}||V_{cd}|)$ as usual 
defines one side of the unitarity triangle. We take $R_b=0.39\pm 0.04$.

Strong phases originate from both $T$ and $P$; 
thus we have $T = |T| e^{i\delta_T}$ and $P = |P| e^{i\delta_P}$  and 
\begin{equation}
a_{\rm CP}^{\rm dir} = \frac{-2 |R|\,  \sin(\delta_P - \delta_T)\, \sin \gamma}{1 -2 |R| \, \cos(\delta_P - \delta_T)\, \cos \gamma + |R|^2}\,.
\end{equation}
In the previous sections we have calculated the contributions
arising from the penguin contractions of the operators ${\cal O}_1^c$  
and ${\cal O}^u_1$.
In addition to these we have also to take into account the tree and
penguin contributions
of the penguin operators ${\cal O}_{3-6}$  and the gluonic penguin 
contribution of the dipole operator ${\cal O}_{8g}$.  
For the latter we use the LCSR result from \cite{KMU}.  The electroweak penguin 
contributions to $B \to \pi \pi$ are color-suppressed and very likely
negligible.

\begin{figure}
\begin{center}
\includegraphics*[width=9cm]{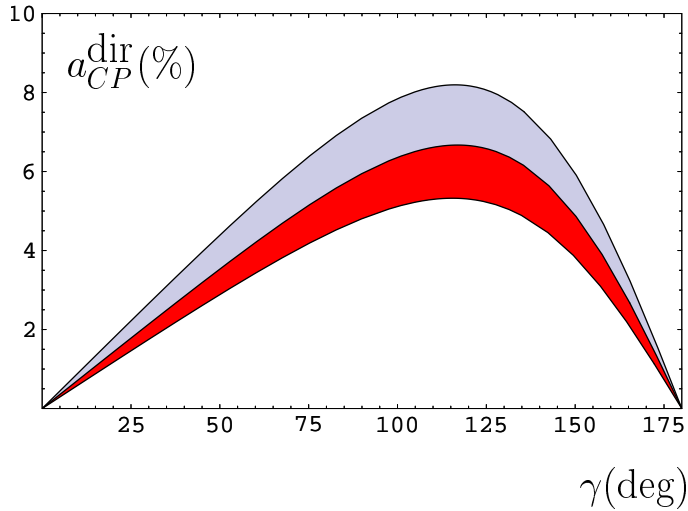} 
\caption{{\em Direct CP asymmetry in $B\to \pi\pi$ as a function of the CKM
  angle $\gamma$. The upper curve is the result obtained for 
$m_b \to \infty$. The dark region is the LCSR result, with all uncertainties from the method included 
(uncertainties in the CKM matrix elements are not taken into
account). The light region  shows the deviation from the $m_b \to \infty$ 
limit result. }
}
\label{fig:CP}
\end{center}
\end{figure}

Apart from the penguin contractions one also has to include the hard
nonfactorizable $O(\alpha_s)$ and soft $1/m_b$ corrections (in the emission topology) to
both $T$ and $P$, obtained, respectively
in the QCD factorization approach \cite{BBNS} and from LCSR
\cite{AK}. It turns out that their contributions to the CP-asymmetry are negligibly small. 
To this end, we can numerically study $a_{\rm CP}^{\rm dir}$  using the penguin
contributions calculated from LCSR at finite $m_b$ 
and compare the result to the infinite-mass limit that agrees with
the QCD factorization prediction \cite{BBNS}. 
The result for $a_{\rm CP}^{\rm dir}$ as a function of $\gamma$ 
is shown in Fig.~\ref{fig:CP}. In this calculation the Wilson coefficients
in $H_{eff}$ are taken 
at the same scale $\mu_b\simeq m_b/2$ as used in LCSR.  
Let us emphasize that this is not yet the final prediction because the 
annihilation effects in $B\to \pi\pi$ are not  included in this analysis. 
The main aim of this numerical illustration is to demonstrate that there
is a  
difference between the finite $m_b$ and the $m_b\to \infty$ result
and that this difference could be sizable (within the uncertainty of
the LCSR prediction).

{\bf 7.} Concluding, we estimated the $B\to \pi\pi$ hadronic matrix elements 
with penguin topology for the current-current operators with 
$c$ and $u$ quarks using the LCSR approach. The main contribution to
the sum rule stems from the $O(\alpha_s)$  quark loop annihilating to a
hard gluon. This contribution determines 
the strong phase of the hadronic matrix element 
justifying the use of the (perturbative) BSS mechanism. 
The soft-gluon effects, which in the sum rule approach 
correspond to multiparticle pion DA's, are suppressed, at least by
$O(\alpha_s/m_b^2)$.  
In $m_b\to \infty$ limit our result 
agrees with the QCD factorization prediction for the penguin contractions.
In both approaches a small value of the direct CP asymmetry in
$B_d^0\to \pi^+\pi^-$ is expected, due to the fact 
that the strong phase is generated perturbatively. 
At finite $m_b$ we predict an even smaller numerical
value for this effect, indicating that 
$O(\alpha_s/m_b)$ corrections are important. Our result does not support 
models employing hadronic dispersion relations 
with intermediate charmed meson pairs, 
predicting sizable charming-penguin effects \cite{models}. 
The method and the results of this paper can be used for analysing
penguin effects in other charmless two-body $B$ decays.
The remaining problem for these decays is the estimate 
of the annihilation effects which is however a very challenging 
and technically difficult task even for the LCSR method.

\section*{Acknowledgment}
We are grateful to M.~Beneke and M.~Shifman for useful discussions.
This work is supported by the DFG Forschergruppe 
"Quantenfeldtheorie, Computeralgebra und Monte Carlo 
Simulationen", and by the German Ministry for Education and Research (BMBF).
B.M. would like to acknowledge hospitality of 
the theory group of Universit{\"a}t W{\"u}rzburg 
and partial support by the Alexander 
von Humboldt Foundation and the Ministry of Science and Technology 
of the Republic of 
Croatia under the contract 0098002.

\end{document}